\documentclass[traditabstract]{aa}

\usepackage{txfonts}
\usepackage{natbib}
\usepackage{graphicx}
\usepackage{times}
\bibpunct{(}{)}{;}{a}{}{,}
\newcommand{\vo}{V2368~Oph }
\newcommand{\ve}{V2368~Oph}
\newcommand{\spefo}{{\tt SPEFO} }
\newcommand{\spefoe}{{\tt SPEFO}}
\newcommand{\phoebe}{{\tt PHOEBE} }
\newcommand{\phoebee}{{\tt PHOEBE}}
\newcommand{\fotel}{{\tt FOTEL} }
\newcommand{\fotele}{{\tt FOTEL}}
\newcommand{\korel}{{\tt KOREL} }
\newcommand{\korele}{{\tt KOREL}}

\newcommand{\ubv}{\hbox{$U\!B{}V$}}
\newcommand{\bv}{\hbox{$B\!-\!V$}}
\newcommand{\ub}{\hbox{$U\!-\!B$}}
\newcommand{\uvby}{\hbox{$uvby$}}
\newcommand{\hp}{\hbox{$H_{\rm p}$}}
\newcommand{\oc}{\hbox{$O\!-\!C$}}

\newcommand{\p}{$\pm$}

\newcommand{\m}{\ifmmode^{\rm m}\!\!.\else$^{\rm m}\!\!.$\fi}
\newcommand{\D}{$^{\rm d}\!\!.$}

\newcommand{\kms}{km~s$^{-1}$ }
\newcommand{\ks}{km~s$^{-1}$}

\newcommand{\tef}{$T_{\rm eff}$ }
\newcommand{\teff}{$T_{\rm eff}$}
\newcommand{\lgg}{{\rm log}~$g$ }

\newcommand{\ANG}{\accent'27A}

\newcommand{\Ame}{\ANG~mm$^{-1}$}
\newcommand{\ha}{H$\alpha$ }

\newcommand{\hae}{H$\alpha$}

\begin{document}

   \title{V2368~Oph: An eclipsing and double-lined spectroscopic binary
used as a photometric comparison star for U~Oph
\thanks{Based on new spectral and photometric observations from
 the following observatories: Dominion Astrophysical Observatory,
 Hvar, Ond\v{r}ejov, San Pedro M\'artir, Tubitak National Observatory, and ASAS service}
\thanks{Tables 2, 3, and 4 are available only in electronic form
 at the CDS via anonymous ftp to cdarc.u-strasbg.fr (130.79.128.5)
 or via http://cdsweb.u-strasbg.fr/cgi-bin/qcat?J/A+A/}
}

\author{P.~Harmanec\inst{1}\and H.~Bo\v{z}i\'c\inst{2}\and P.~Mayer\inst{1}\and
    P.~Eenens\inst{3}\and M.~Bro\v{z}\inst{1}\and M.~Wolf\inst{1}\and
    S.~Yang\inst{4}\and M.~\v{S}lechta\inst{5}\and D.~Ru\v{z}djak\inst{2}\and
    D.~Sudar\inst{2}\and H.~Ak\inst{6}}

   \institute{
    Astronomical Institute of the Charles University,
    Faculty of Mathematics and Physics,\\
    V Hole\v sovi\v ck\'ach 2, CZ-180 00 Praha 8, Czech Republic
\and
    Hvar Observatory, Faculty of Geodesy, University of Zagreb,
    10000 Zagreb, Croatia
\and
    Departamento de Astronomia, Universidad de Guanajuato, Apartado 144,
    36000 Guanajuato, GTO, Mexico
\and
    Physics \& Astronomy Department, University of Victoria,
    PO Box 3055 STN CSC, Victoria, BC, V8W 3P6, Canada
\and
   Astronomical Institute, Academy of Sciences of the Czech Republic,
   CZ-251~65~Ond\v{r}ejov, Czech Republic
\and
   Department of Astronomy and Space Sciences, Faculty of Sciences,
   Erciyes University, 38039 Kayseri, Turkey
    }

   \titlerunning{Eclipsing binary \ve}

   \offprints{Petr Harmanec\\ \email: Petr.Harmanec$@$mff.cuni.cz}

   \date{Release \today}

\abstract{The A-type star HR~6412 = \vo was used by several investigators
as a photometric comparison star for the known eclipsing binary U~Oph but was found
to be variable by three independent groups, including us.
By analysing series of new spectral and photometric observations
and a critical compilation of available radial velocities, we were able
to find the correct period of light and radial-velocity variations and
demonstrate that the object is an eclipsing and double-lined spectroscopic
binary moving in a highly eccentric orbit. We derived a linear ephemeris
$T_{\rm min.I}=$HJD~(2454294.67\p0.01) + (38\fd32712\p0\D00004) $\times E$
and estimated preliminary basic physical properties of the binary.
The dereddened \ubv\ magnitudes and effective temperatures of the primary
and secondary, based on our light- and velocity-curve solutions, led
to distance estimates that agree with the Hipparcos
distance within the errors. We find that the mass ratio must be close
to one, but the limited number and wavelength range of our current spectra
does not allow a truly precise determination of the binary masses.
Nevertheless, our results show convincingly that both binary components are
evolved away from the main sequence, which makes this system astrophysically
very important. There are only a few similarly evolved A-type stars
among known eclipsing binaries.  Future systematic observations and
careful analyses can provide very stringent tests for the stellar
evolutionary theory.
}

\maketitle


\section{Introduction}\label{one}
HR~6412 = HD~156208 has often been used as the photometric comparison star
for a well-known eclipsing binary U~Oph, which exhibits
apsidal motion \citep{hufkop51,koch77,wolf2002,vaz2007}.
\citet{iccd1} reported that HR~6412 is a speckle-interferometric binary
with a separation of 0\farcs136 and estimated the orbital period to 72 years.
However, \citet{iccd10} could not resolve this pair and concluded that
the original detection had been spurious.

In Table~\ref{vmags}, we summarize the various determinations of the yellow
magnitude of \vo published by several authors. It seems to indicate
that no secular variations in its brightness have been recorded, since the
scatter of values of a few hundredths of a magnitude is quite normal
for yellow magnitudes recorded in different photometric systems.

\begin{table}
\caption[]{Published yellow magnitudes of \ve.}\label{vmags}
\begin{flushleft}
\begin{tabular}{lllllc}
\hline\noalign{\smallskip}
\ \ Mag.  & Source & Photometric system \\
\noalign{\smallskip}\hline\hline\noalign{\smallskip}
6\m17 & \protect{\citet{eggen55}} & $(P,V)_{E}$ system \\
6\m19 & \protect{\citet{stokes72}}& \uvby\ and $H\beta$ \\
6\m16 & \protect{\citet{becker75}}& Cousins' values \\
6\m178& \protect{\citet{gro76}}   & \uvby; 1965 -- 1970 \\
6\m177& \protect{\cite{sowell93}} & \uvby; Nov. 1988 \& Apr. 1991\\
6\m18 & \protect{\cite{vangent82}}& $BVR$; 1970 \\
\noalign{\smallskip}\hline
\end{tabular}
\end{flushleft}
\end{table}

During our 2001 observations of U~Oph at San Pedro M\'artir Observatory
(SPM hereafter), HR~6412 was also used as the comparison star.
We noticed large changes in this
comparison on JD~2452071.71--1.85. Upon a literature search, we found that
the variability of HR~6412 has been discovered by \citet{esa97},
who classified it as an eclipsing binary with a period of 7\fd70.
\citet{kazaretal99} then assigned it the variable-star name \ve.
The variability has also been confirmed by \citet{vaz2007}, who
mention that the period found by \citet{esa97} was incorrect but give
no other details.

  The main goal of this study is to publish the first correct and accurate
linear ephemeris of \ve, which can be used to  correct earlier
photometric observations of U~Oph. We also derive preliminary orbital and
light-curve solutions and show that they can lead to self-consistent
basic physical properties of the binary. However, considering the limited
amount and heterogeneity of our observational material, we do not aim
to determine the final, accurate physical elements of the system.


\section{Observational material used}\label{two}
\subsection{Photometry}
After realising that \vo is a variable star, we started systematic
\ubv\ observations of it at Hvar and SPM observatories. A limited set of
\ubv\ observations was also obtained by HA at the Turkish National
Observatory. Besides, we compiled the Hipparcos \hp\ observations
and $V$ photometry from the ASAS project \citep{pojm2002}. Details
on data sets and their reduction  are in Appendix~\ref{aone}, and all
individual \ubv\ and $V$ observations with their HJDs are provided
in Table~2 (electronic only).

\begin{figure}
\resizebox{\hsize}{!}{\includegraphics{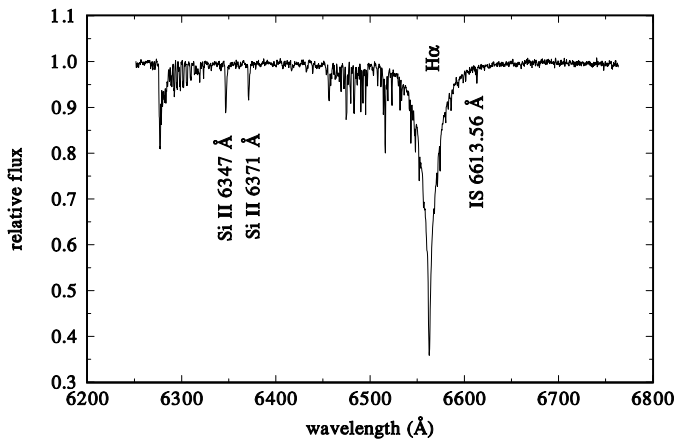}}
\caption{One complete Ond\v{r}ejov red spectrum of \ve, taken on
HJD~2454357.2917, which shows that the only stronger lines, suitable for
the RV measurements, are \ha and the \ion{Si}{II}~6347 and 6371~$\AA$.
The spectrum contains many water vapour and oxygen telluric lines
and the interstellar line at 6613.56~$\AA$.}
\label{red-sp}
\end{figure}

\begin{figure}
\resizebox{\hsize}{!}{\includegraphics{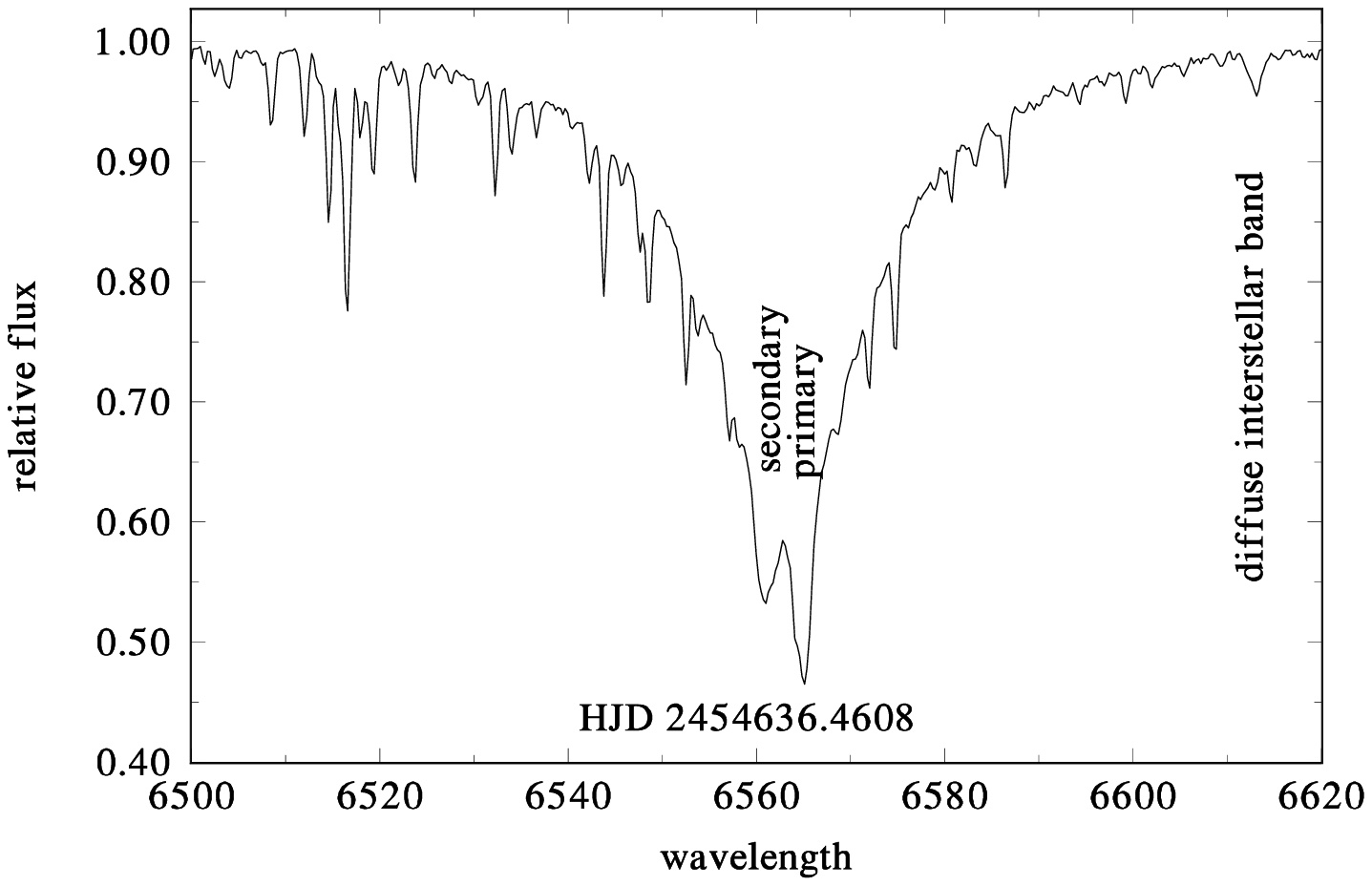}}
\resizebox{\hsize}{!}{\includegraphics{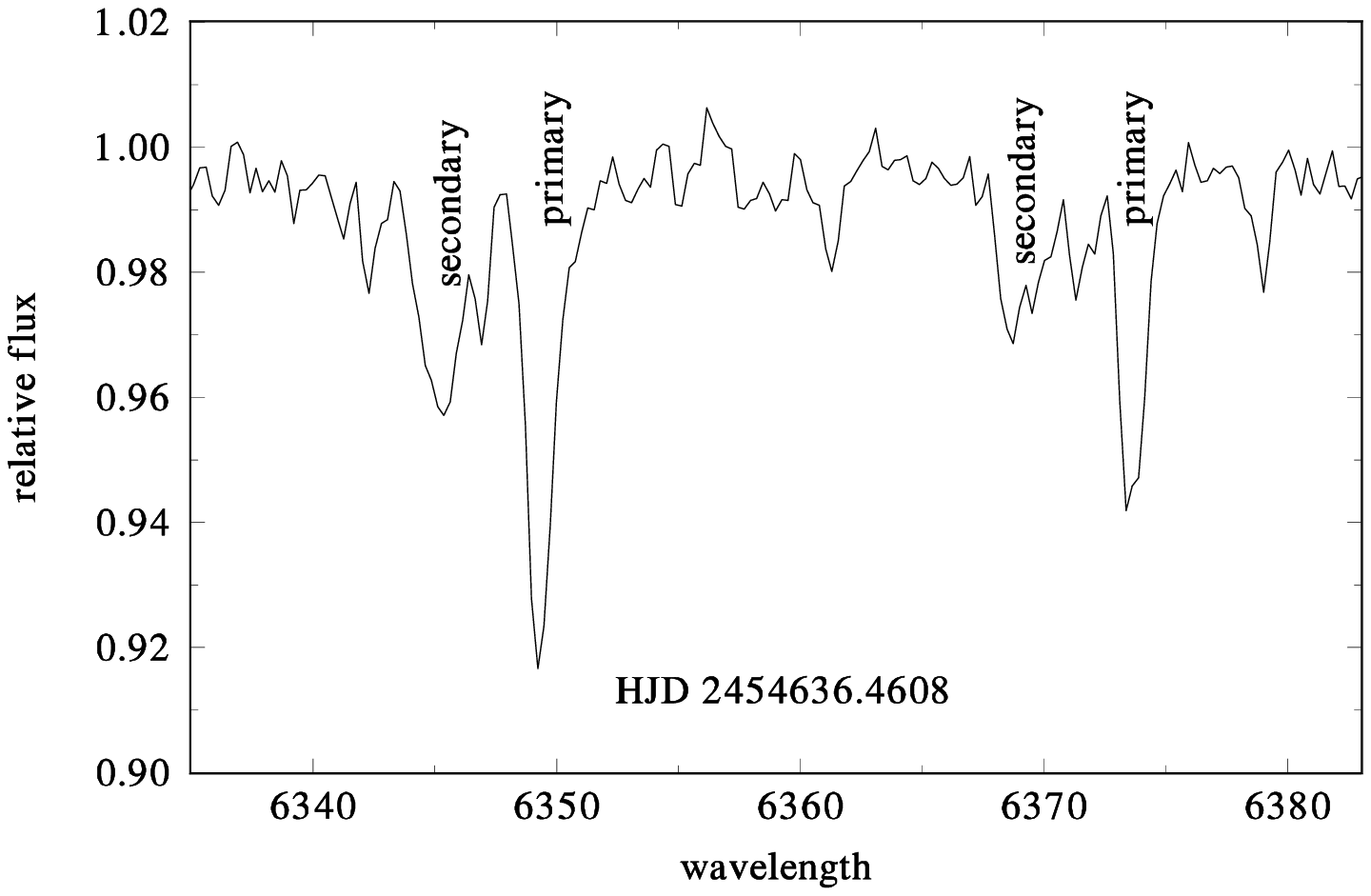}}
\caption{The \ha and \ion{Si}{II}~6347 and 6371~\ANG\ line profiles
from the Ond\v{r}ejov spectrum taken near one periastron passage
on HJD~2454636.4608. One can clearly see the lines of both binary
components, separated by more than 200 \kms in their radial velocity.}
\label{profs}
\end{figure}


\subsection{Spectroscopy}
Simultaneously with photometric observations, we also begun to collect
electronic spectra in Ond\v{r}ejov, San Pedro M\'artir, and
Dominion Astrophysical Observatory (OND, SPM, and DAO hereafter).
A detailed discussion of all spectra, their reduction, and a journal of
observations can be found in Appendix~\ref{btwo}.

 Here, we only want to add a few comments relevant to further analyses.
The only spectral region that is available in the spectra from all three
observatories is the red region containing only three strong enough
spectral lines suitable for the RV measurements: the Balmer \ha line
and the doublet of \ion{Si}{II}~6347 and 6371~\ANG\ lines (see Figs.~\ref{red-sp} and \ref{profs}).
The SPM spectra also cover the region of \ion{Mg}{II}~4481~$\AA$ line,
in which both components are clearly seen, so this line was also found suitable
for the RV measurements. The RV measurements were carried out in three
different ways.
(1)~We used the program \spefo \citep{sef0,spefo},
which permits the RV measurements via sliding direct
and flipped line profiles on the computer screen until a perfect match is obtained.
(2)~We fitted the observed line profiles by two Gaussians
shifted in their positions in such a way as to obtain the best match of
the observed, often blended line profiles of the primary and secondary.
(3)~We verified our RV values by an automated fitting of a combination
of two synthetic spectra --- selected from the Ond\v rejov library of synthetic spectra,
prepared and freely distributed by Dr. J. Kub\'at ---
to each of the observed spectra using a simplex algorithm.
The $\chi^2$ \citep[defined in][]{brozetal2010} was calculated for
the {\em entire\/} red spectrum in the wavelength range
$6256\hbox{ to }6768\,\AA$,
which includes all the individual lines analysed previously.

The second method is preferred as it returns the most likely velocity
amplitudes. The results of the third method are statistically compatible.
The first method (\spefo RV measurements) usually
underestimates the true semi-amplitudes of the orbital motion,
because of the line blending, especially for the steep wings of \hae.
We, therefore, used the \spefo\ \ha RVs only for the initial search
of the orbital period, to combine them with older published RVs,
measured in a standard way from the photographic spectra.

For all red spectra, we followed the procedure outlined
in \citet{sef0} and measured RVs of a selection of unblended
{\em telluric\/} lines in \spefoe. We then used the difference between
the calculated heliocentric RV correction and the true mean RV of telluric
lines to correct the zero point of the RV scale individually for each
spectrogram. Regrettably, these corrections were less reliable for
the majority of the red SPM spectra, which only contain very weak telluric
lines owing to the high altitude of that observatory. No such corrections
were possible for the blue SPM spectra in the absence of telluric lines,
of course.

Individual \hae, \ion{Si}{II}, and \ion{Mg}{II} RVs,
measured in a standard way in \spefo and via Gaussian fits
to line profiles with their HJDs of mid-exposures are in
(electronic only) Table~3. The rectified and wavelength-calibrated
spectra are in (also electronic) Table~4.

\setcounter{table}{4}

\begin{figure}
\resizebox{\hsize}{!}{\includegraphics{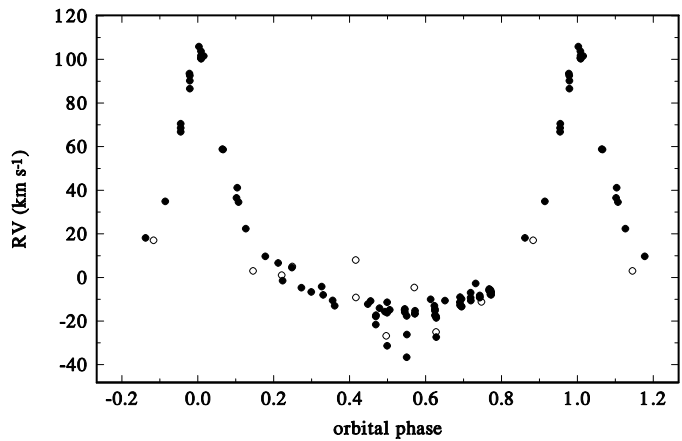}}
\caption{The radial-velocity curve of the primary component of \vo
based on \ha RVs measured in \spefo for our OND, DAO, and SPM spectra
(filled circles) and published RVs (open circles) for the period of
38\fd3307 from the \fotel orbital solution with phase zero at
periastron passage. See the text for details.}
\label{rv1}
\end{figure}


\section{Preliminary analysis and search for a correct orbital period}
As our observations progressed, it soon became obvious that the orbital
period must be much longer than the 7\fd70 period reported by \citet{esa97}
and that the orbit had to have a high eccentricity, since we were observing
a constant brightness and only small RV changes. When we finally
succeeded in observing a decline into the minimum on the night
JD~2454294.35-4.54, we were able to combine it with earlier
minima recorded by Hipparcos, ASAS, and our discovery observation
at SPM and to obtain the first guess that the period should be close
to 38 days. Continuing spectroscopic observations then allowed us
to cover parts of two periastron passages on JD~2454366-67, and
JD~2454636 and an iterative analyses of the RV and light curves
allowed us to estimate the value of the orbital period to 38\fd33.

There are two limited sets of earlier RV measurements. \citet{christie26}
published five RVs covering the interval JD~2423233.8--995.8, and
\cite{rob135} published another five RVs from low-dispersion spectra
covering JD~2437441.6-740.7. We combined these RVs with our own
RV measurements in the program \spefo \citep{sef0, spefo} for the \ha line and used the \fotel program
\citep{fotel1, fotel2} to derive preliminary orbital elements
and a more accurate value of the period.
We obtained $P = 38\fd3302\pm0\fd0015$,
$T_{\rm periast.} = $ HJD$~54290.894\pm0.096$,
$T_{\rm min\,I} = $ HJD~54294.417,
$e=0.551\pm0.010$, $\omega=355.2\pm1.6$,
$K_1=59.71\pm4.1$~\ks,
$\gamma_{\rm old}=17.0\pm3.4$, and
$\gamma_{\rm new}=10.07\pm0.50$, the rms errors of the model fit
to the data per 1 observation being
10.3, and 4.1~\kms for the old and new RVs.

\begin{figure}
\resizebox{\hsize}{!}{\includegraphics{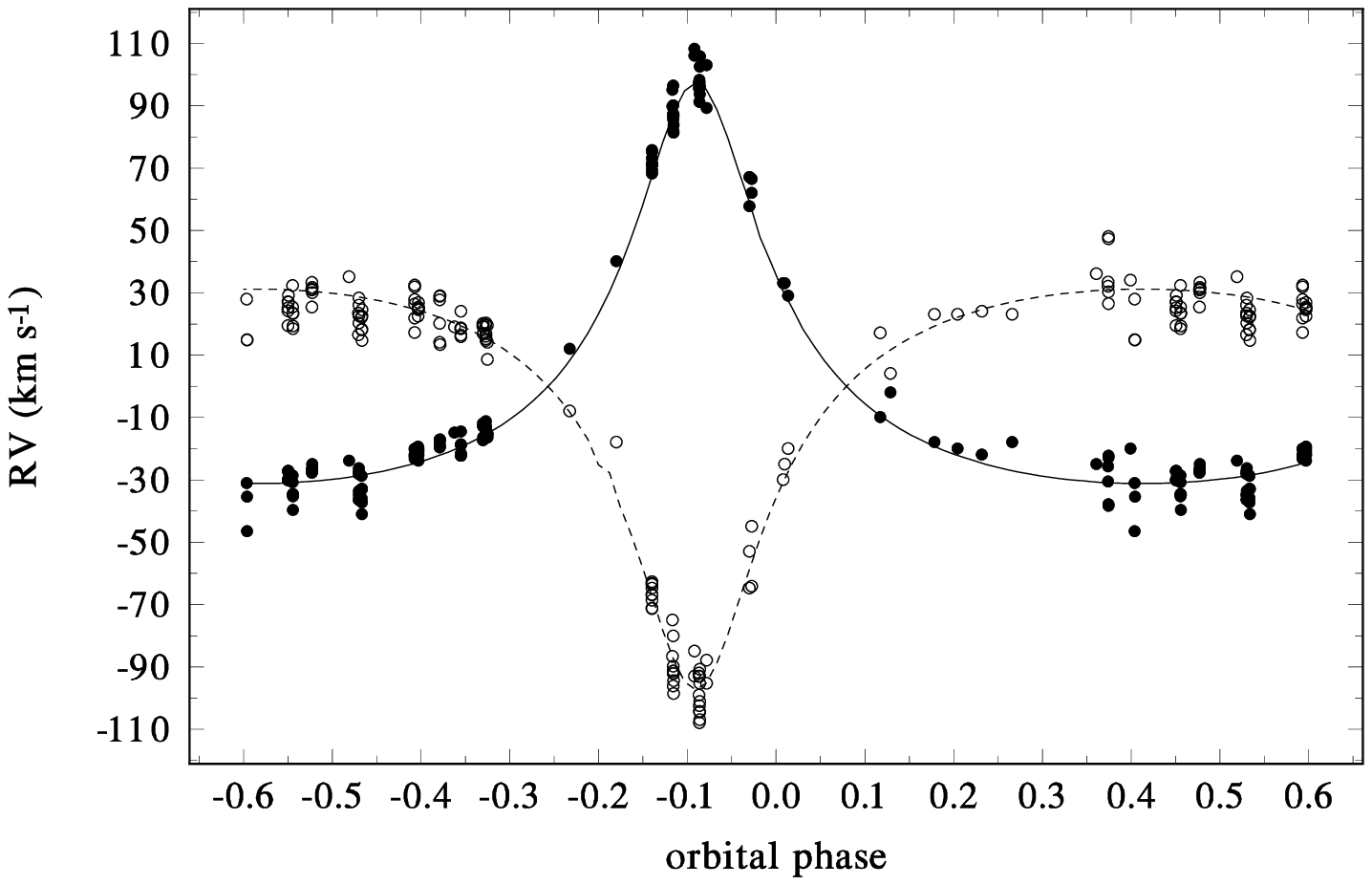}}
\resizebox{\hsize}{!}{\includegraphics{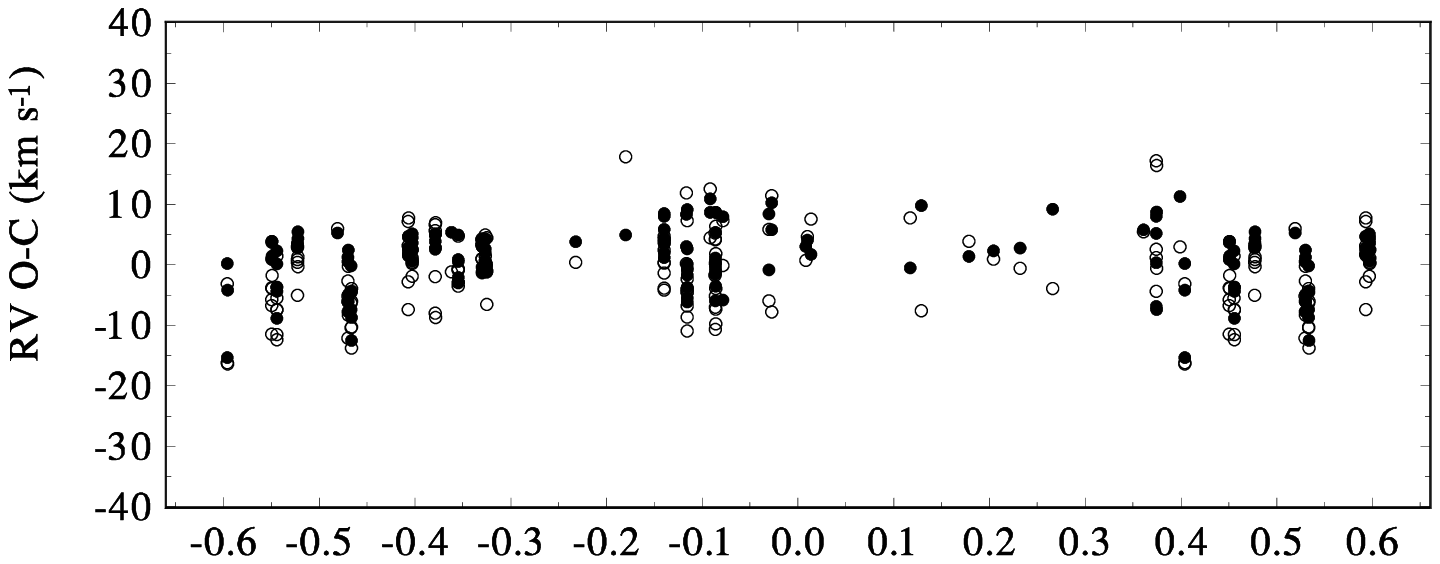}}
\caption{The final radial-velocity curves of both binary
components of \vo based on Gaussian fits (circles) and the theoretical RV curves
based on our \phoebe solution (lines). The lower panel shows the \oc\
residua from the solution. Typical observational uncertainties
are 10~\ks. Orbital phases from ephemeris(\ref{efe}) are used.}
\label{rvc}
\end{figure}

\begin{figure}
\resizebox{\hsize}{!}{\includegraphics{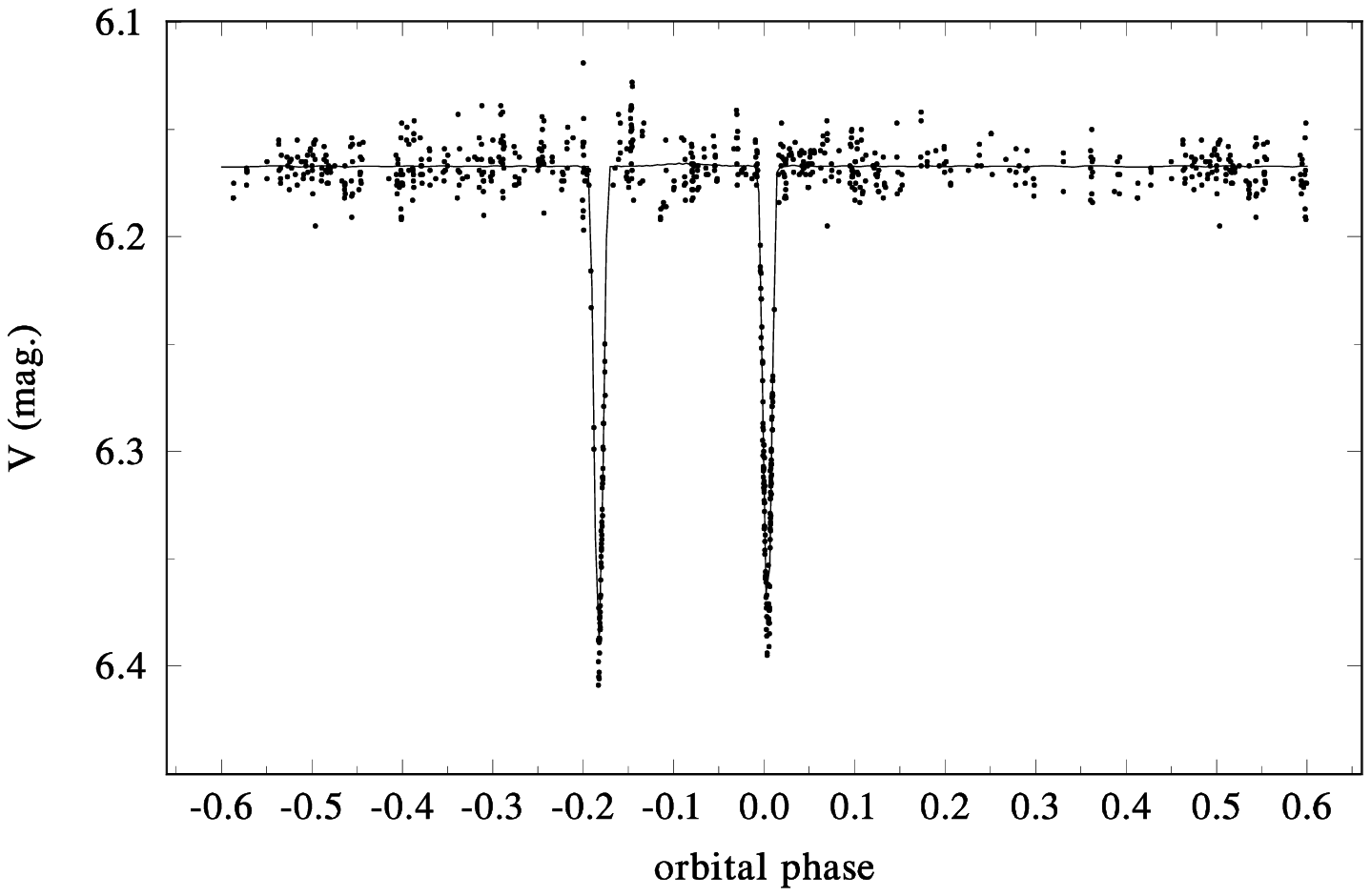}}
\resizebox{\hsize}{!}{\includegraphics{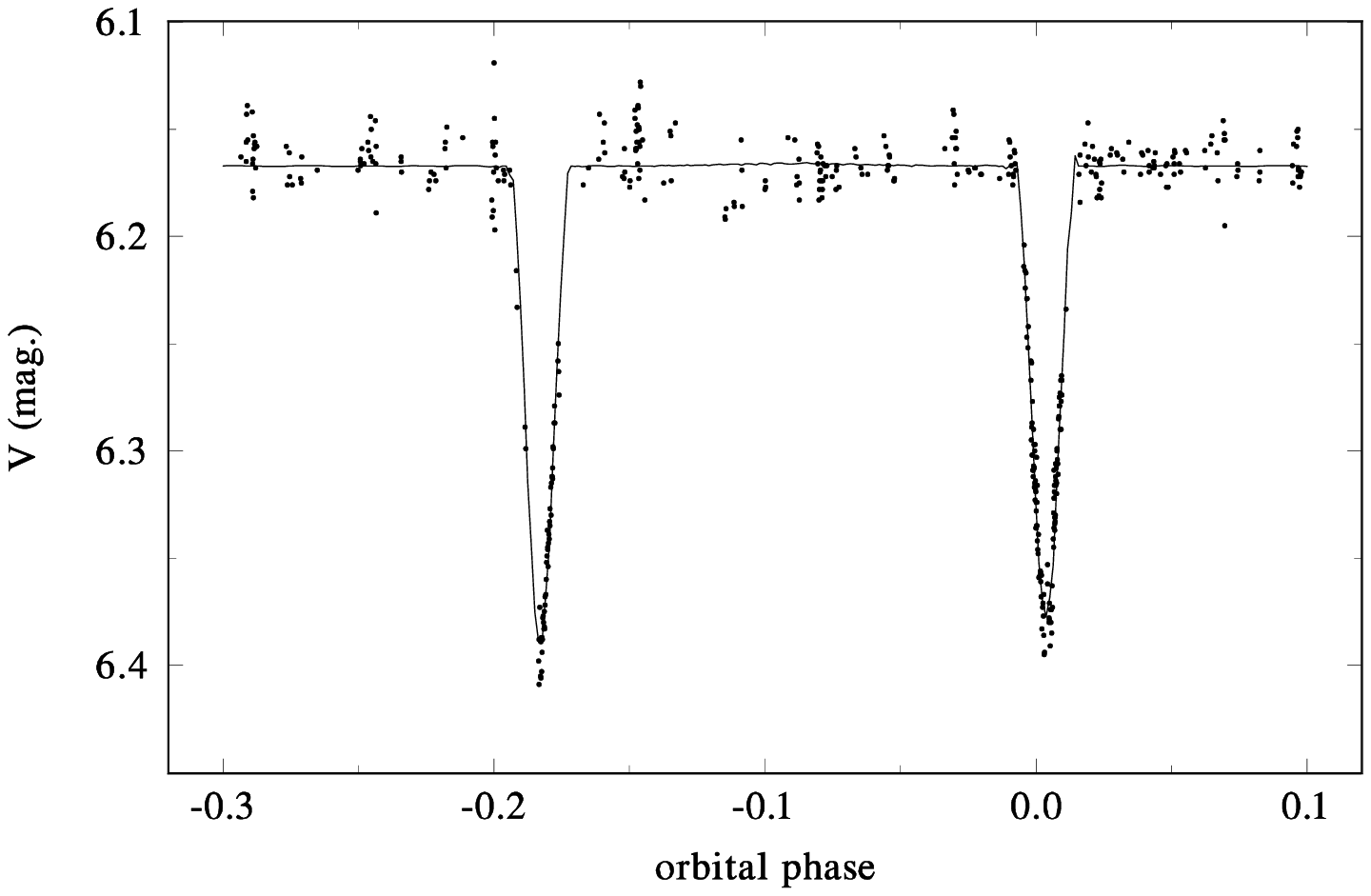}}
\resizebox{\hsize}{!}{\includegraphics{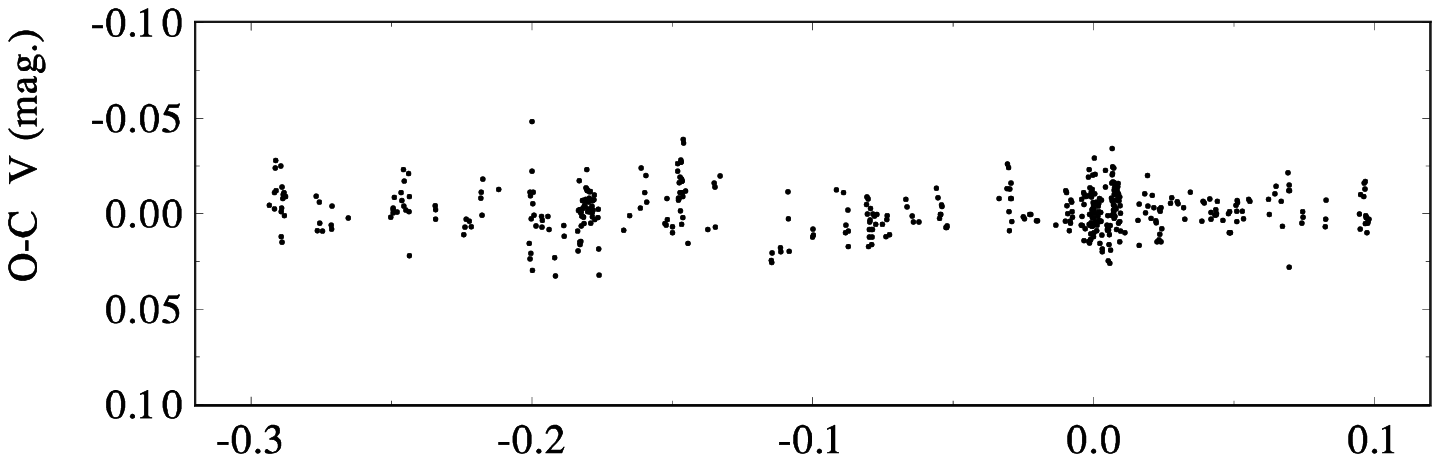}}
\caption{The observed $V$-band light curve compared to the theoretical one,
based on our \phoebe solution. The lower panels show a zoom of the curves
in the neighbourhood of both binary eclipses and the \oc\ residua
from the model fit. Typical 1-$\sigma$ observational uncertainties are 0\m01.
Orbital phases from ephemeris (\ref{efe}) are used.}
\label{vlc}
\end{figure}

\begin{table*}
\caption[]{Published \uvby\ and H$\beta$ observations of \ve.}\label{uvby}
\begin{flushleft}
\begin{tabular}{llllclcl}
\hline\noalign{\smallskip}
\ \ $V$  & $b-y$   &$m_1$    &$c_1$    &N$_{\uvby}$&H$\beta$&N$_{\rm H\beta}$&Source\\
\noalign{\smallskip}\hline\hline\noalign{\smallskip}
 --      &0\m167   &0\m090   &1\m195   &5&2.850   &4&\protect{\citet{craw72}}\\
6\m19    &0\m177   &0\m095   &1\m216   &3&2.860   &3&\protect{\citet{stokes72}}\\
6\m178(1)&0\m161(2)&0\m095(2)&1\m204(2)&2& --     & &\protect{\citet{gro76}}\\
 --      &  --     &  --     &  --     & &2.870(2)&3&\protect{\citet{gro77}}\\
6\m177(5)&0\m189(2)&0\m072(5)&1\m203(5)&3& --     & &\protect{\cite{sowell93}}\\
\noalign{\smallskip}\hline
\end{tabular}
\end{flushleft}
\end{table*}

\begin{table*}
\begin{center}
\vspace{0.1cm} \caption[ ]{The final combined light-curve and
RV-curve solutions obtained with \phoebee.}
\label{fotrvsol}
\begin{tabular}{lc|lcl|lcl}
\hline\hline\noalign{\smallskip}
Element &                & Primary & Binary  & Secondary& Primary & Binary  & Secondary \\
\noalign{\smallskip}\hline\noalign{\smallskip}
$P$                  &(d)&& 38.327115(43) &      && 38.327118(43) & \\
$T_{\rm periastr.}$&(RJD)&& 54291.039(11) &      && 54291.042(11) & \\
$T_{\rm min.I}$    &(RJD)&& 54294.670(17) &      && 54294.668(17) & \\
$T_{\rm min.II}$   &(RJD)&& 54287.498(17) &      && 54287.495(17) & \\
$e$  &                   && 0.51527(14) &        && 0.51524(14) & \\
$\omega$  &$({}^\circ)$  && 359.33(20) &         && 359.41(20) & \\
$i$       &$({}^\circ)$  && 86.165(22) &         && 86.139(22) &\\
$r$                      &&  0.0473&&0.0460      &  0.0477&&0.0464\\
$\Omega$  &              & 23.29(11) && 24.75(12)& 23.05(11) && 23.71(11)  \\
$a$       &$(R_{\sun})$  && 83.67(67) &          && 83.70(68) & \\
$K_\mathrm{j}$ &(\ks)    & 65.7(8)&& 62.9(9)     &64.3(6) && 64.3(6) \\
$K_2/K_1$     &          && 1.044(15) &&         & 1.0 fixed &\\
$T_\mathrm{eff}$  &(K)   & 9300 fixed  &&9500(200)& 9300 fixed  && 9500(200)\\
$M$   &($M_{\sun}$)      & 2.62(2)&& 2.74(7)     &2.68(8)&&2.68(8) \\
$R$   &($R_{\sun}$)      &3.96(2) && 3.84(2)     &3.99(3)&&3.87(2) \\
$M_\mathrm{bol}$ &(mag)  & $-0.31(9)$ && $-0.34(9)$      & $-0.33(9)$ &&$-0.36(9)$ \\
$\lgg$     &[cgs]        & 3.66(1) && 3.71(1)    & 3.66(3) && 3.69(3) \\
$L_\mathrm{j}$&$V$ band  & 0.5029(30)  && 0.4971 & 0.5024(30)  && 0.4976  \\
$L_\mathrm{j}$&$B$ band  & 0.5004(32)  && 0.4996 & 0.4995(32)  && 0.5005  \\
$L_\mathrm{j}$&$U$ band  & 0.4942(37)  && 0.5058 & 0.4927(38)  && 0.5073  \\
$V$       &(mag.)        &6.913(16)& 6.167(11) &6.926(16) &6.913(16)& 6.166(11)&6.924(16)\\
$B$       &(mag.)        &7.137(17)& 6.386(11) &7.139(17) &7.139(17)& 6.385(11) &7.137(17)\\
$U$       &(mag.)        &7.363(20)& 6.598(13) &7.338(20) &7.367(20)& 6.599(13) &7.335(20)\\
$V_0$     & (mag.)       & 6.26(8) && 6.29(8)  & 6.26(8) && 6.29(8)  \\
$(\bv)_0$ & (mag.)       & 0.021(11)&& 0.014(11)& 0.021(11) && 0.014(11)\\
$(\ub)_0$ & (mag.)       & 0.077(15)&& 0.054(15)& 0.078(15)&& 0.053(15)\\
No. of obs. &  \ubv\ / RV&&1913 / 268 &&             &1913 /268  & \\
$\chi^2$ &          &&  1726 &&             &1730   & \\
\noalign{\smallskip}\hline\noalign{\smallskip}
\end{tabular}
\tablefoot{Columns 2--4 contain a solution
based on the free convergence; columns 5--7 contain the solution
for the fixed mass ratio of 1.  All epochs are in
${\rm RJD} = {\rm HJD} - 2400000$. Probable elements and their formal error estimates
are provided, where $\Omega$ is the value of the Roche-model potential used
in the WD program, and $L_j$ ($j=1,2$) are the relative luminosities
of the components in individual photometric passbands. They are normalized
so that $L_1+L_2=1$. The number of observations represents
the sum of RVs of the primary and secondary and a sum of individual
observations in all passbands.}
\end{center}
\end{table*}

Since the narrow and steep photometric eclipses are very sensitive to the
phase shifts, we used an interactive computer program (written by HB), which
allows the user to display the phase diagrams based on the observed data
in the neighbourhood of the eclipses for various smoothly varied values
of the orbital period. This way we found that the true orbital period
must be very close to the value of 38\fd3272.


\section{Towards basic physical properties of the binary}
To obtain self-consistent physical properties of the components and
the binary system, we had to proceed in an iterative way.
 We selected several stronger lines seen in both binary components and derived
their RVs via Gaussian fits to line profiles. In particular, we used
the \ion{Si}{II}~6347\&6371~\ANG\ and \ha lines, available in all spectra,
and \ion{Mg}{II}~4481~\ANG, measurable in the SPM spectra.
For \hae, the Gaussian profiles were not optimal so we
tentatively disentangled the \ha profiles, using the \korel program
\citep{korel1,korel2,korel3,korel4} and used the disentangled profiles instead of
Gaussians.\footnote{It would seem logical to derive the orbital
solution directly with \korele. However, due to heterogeneity of the available
spectra, their different spectral resolutions and relatively limited number,
this procedure was not satisfactory in the given case.}

We alternatively used the programs \phoebe \citep{prsa05,prsa06} based
on the \citet{wd71} program, and \fotele, already used in the first step,
to derive preliminary values of some critical parameters. To obtain
the best possible estimate of the RV semi-amplitudes, we allowed for
individual systemic velocities for each of the ions used. In particular,
we found the systemic velocities of 2.97\p0.55, 0.14\p0.54, and
9.9\p6.0 for \ion{Si}{II}, \ion{Mg}{II}, and \hae, respectively.
The differences between these values are probably insignificant considering
their errors and the inability to check the zero point of the RV scale
for the blue spectra via measurements of the telluric lines.
Since \phoebe can treat only one joint systemic velocity,
we subtracted the values of respective systemic velocities from the observed
RVs and used these shifted RVs from all three ions as one dataset for
the primary and another one for the secondary in \phoebee. We then naturally
kept the systemic velocity fixed at zero in \phoebe solutions.

In the latest (development) version of \phoebe
that we are using, the convergence is governed by minimization
of a cost function $\chi^2$ defined in the case of our datasets as
\begin{equation} \label{cost}
\chi^2 = \sum_p \frac 1{\sigma_p^2} \sum_{i=1}^{N_p}w_i (f_i - s_i)^2,
\end{equation}
where index $p$ denotes the individual photometric passbands,
$\sigma_p$ their standard deviations per 1 observation,
$N_p$ is the number of individual observations for $p$-th passband,
$w_i$ are standard weights of individual observations, and $f_i$ and $s_i$
the observed and calculated fluxes, respectively. The value of
the $\chi^2$ function is tabulated along with the solutions.

Although it should be possible to derive the effective temperatures
of both binary components from calibrated \ubv\ photometry
\citep{prsa06, wilson2008}, the propagation of errors often leads
to unreliable results. For that reason we restricted ourselves to
the standard approach of estimating the effective temperature
of the primary from the dereddened colours and observed spectra,
then keeping its value fixed in the solutions.

There are four sets of \uvby\ observations and three sets of H$\beta$
photometry of \vo -- see Table~\ref{uvby}.
Using the program {\tt UVBYBETA} written by T.T.~Moon and modified
by R.~Napiwotzki, which is based on a calibration devised by \citet{moon1985},
we found that those sets of Str\"omgren photometry imply
a mean effective temperature of the binary between 8900~K, and 9400~K
and a mean \lgg between 3.50 and 3.61.

Since both stars are detached well even near periastron, the light-curve
solution basically does not depend on the exact value of the mass ratio.
Considering this, a preliminary \phoebe solution was used to derive relative
luminosities of both components, \ubv\ magnitudes of the binary at
maximum light from the \ubv\ observations transformed to the standard
system and from them the \ubv\ magnitudes of the primary and secondary in
each passband. These were then dereddened in a standard way, assuming
$A_V=3.2\,E($\bv). The dereddened magnitudes
and indices confirmed the spectral type of A2 for the primary.
Using \cite{flower1996} calibration of (\bv)$_0$
indices vs. bolometric corrections and \teff, we estimated the effective
temperature of the primary to (9300\p200)~K.
Keeping the value of 9300~K fixed, we derived a freely converged \phoebe\
solution, which is presented in detail in Table~\ref{fotrvsol}. The rms
errors of all converged parameters in \phoebe are derived from a covariance
matrix. For other, deduced parameters, we propagated the errors to obtain
the estimates given in the Table.  Since a realistic error of the effective
temperature of the primary is about $\pm200$~K, this must imply
that the formal error of the effective temperature of the secondary,
estimated from a covariance matrix in \phoebee, is too low and must
also be about $\pm200$~K.

It is clear at first sight that this solution is not satisfactory since
it leads to a model in which the more massive component is the less evolved
of the two. We believe that the problem lies in the limited quality of
our radial velocities. As a matter of fact, the solutions for RVs
of individual ions oscillated between 0.95 and 1.05 in the resulting
mass ratio. At the same time, since both binary components are well detached,
the light curve solution is stable and basically does not depend on the
mass ratio. Therefore, we derived another \phoebe solution, this time
with the mass ratio fixed to 1.0. This solution is also provided in the
last three columns of Table~\ref{fotrvsol} and we take it as the reference
solution for the following discussion. The corresponding RV curves and
the light curve in the $V$ band are shown in Figs.~\ref{rvc}, and \ref{vlc},
respectively. It is seen that there is little difference in all parameters,
which do not depend on the mass ratio between the two solutions.
For completeness, we also derived another solution for the mass ratio
of 0.95. This led to a slightly worse $\chi^2 = 1770$, but the
photometric elements were again very similar to those two
shown in Table~\ref{fotrvsol}.

The solutions led to the following linear ephemeris, which should
enable correction of existing photometric observations of U~Oph secured
differentially relative to \ve:

\begin{equation}
T_{\rm min.I}={\rm HJD}\,\,2454294.67 + 38\fd32712\times E\,\,.\label{efe}
\end{equation}

It is encouraging to note that a separate dereddening of
the \ubv\ magnitudes of the primary and secondary led invariably
to $E(\bv)=0\m20$ and to distance moduli of 6.46(12), and 6.47(12) for
the primary and secondary, respectively. The dereddened values of
the secondary indicate a slightly earlier spectral type, in accordance
with its higher effective temperature obtained from the \phoebe solution.
We note that $E(b-y)$ derived with the program {\tt UVBYBETA} from
the \uvby\ values of Table~\ref{uvby} is 0\m15, which agrees well with
the $E(\bv)$ derived by us.\footnote{Note that $E(b-y) = 0.74E(\bv)$.}

The parallax of \vo was obtained by the ESA Hipparcos mission, and its
originally published value \citep{esa97} is 0\farcs00554 \p 0\farcs00086,
while an improved value obtained by \citet{leeuw2007a,leeuw2007b}
reads as 0\farcs00455 \p 0\farcs00048.   The distance modulus
obtained from our photometric solution implies a parallax
of 0\farcs00506, in excellent agreement with the above values,
deduced from the Hipparcos observations.


\section{Stellar evolution of the components}
Since \vo is a detached binary with no evidence of mass transfer, one
can use a one-dimensional program for stellar evolution to see whether
the observed properties of the binary agree with the model prediction.
To this end, we used the stellar-evolution module {\tt MESAstar}
by \citet{pax2011}.

\begin{figure*}
\centering
\includegraphics[width=8cm]{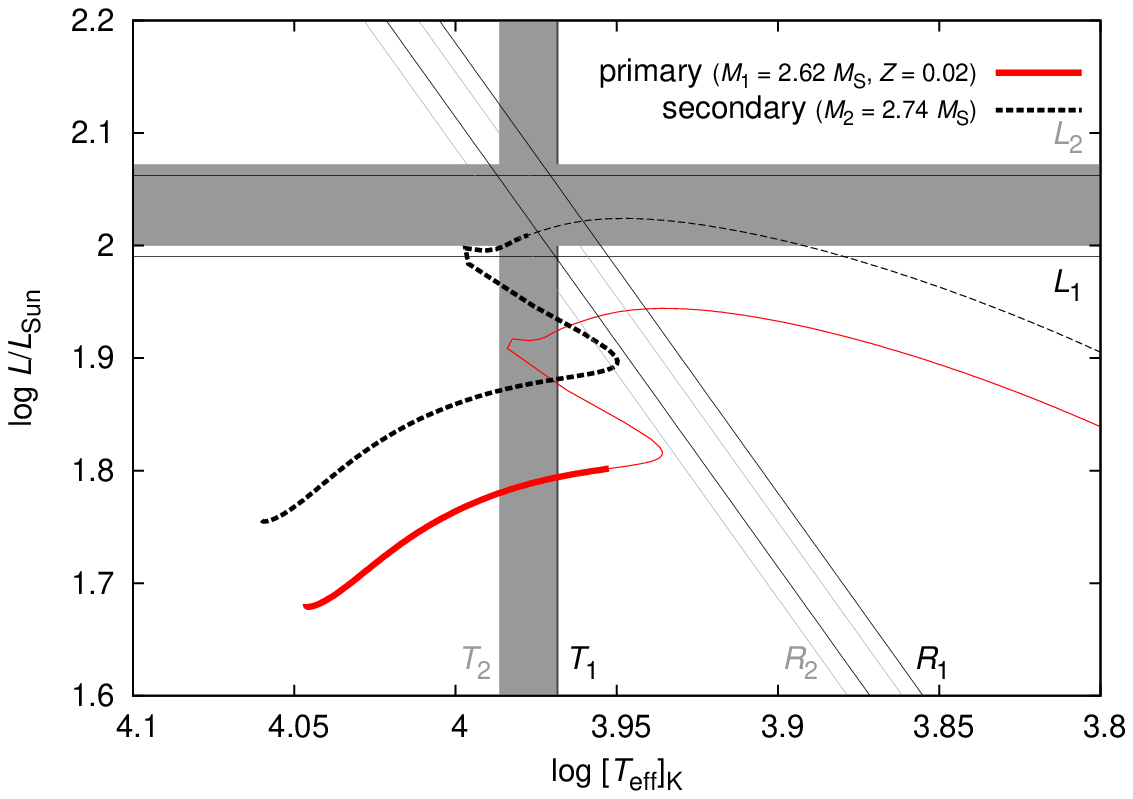}
\includegraphics[width=8cm]{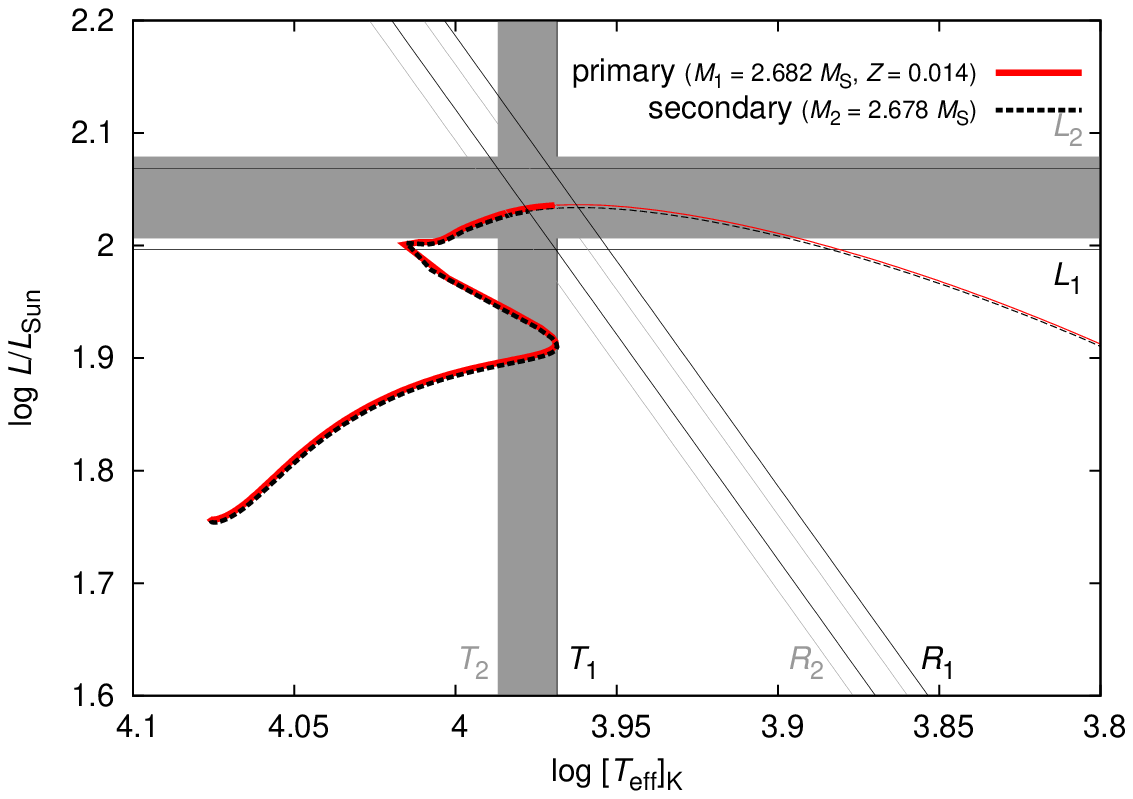}
\includegraphics[width=8cm]{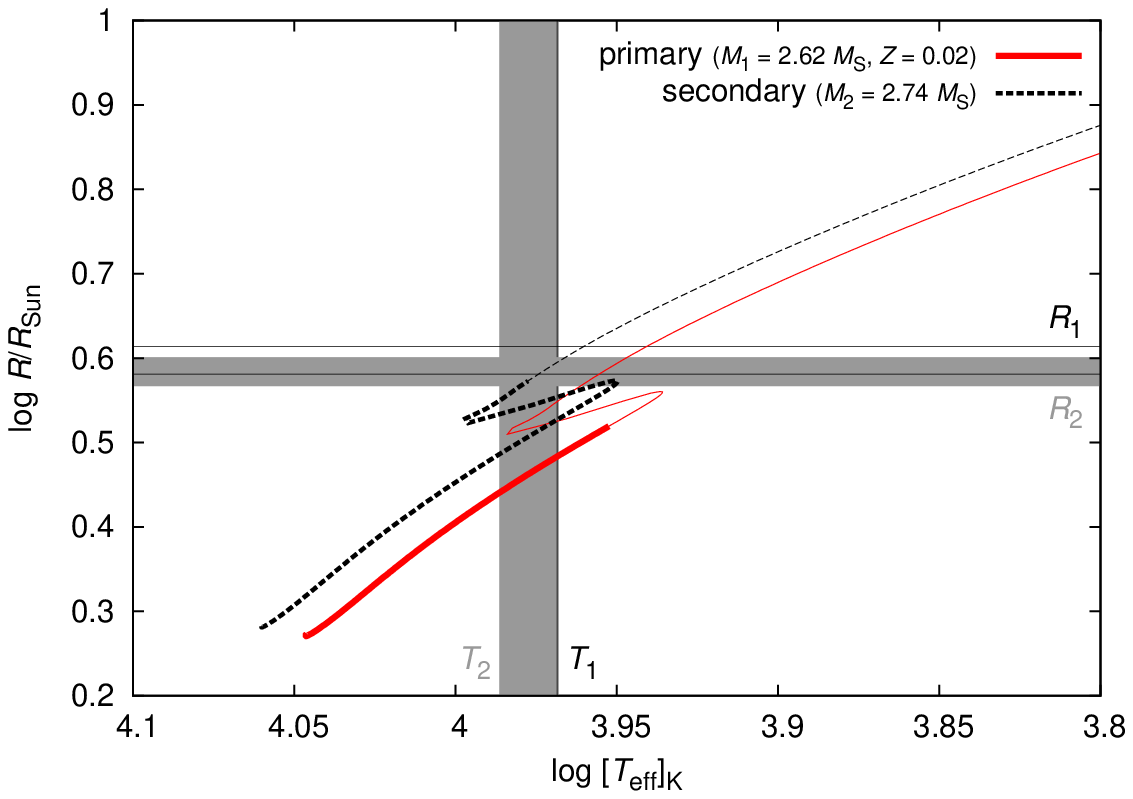}
\includegraphics[width=8cm]{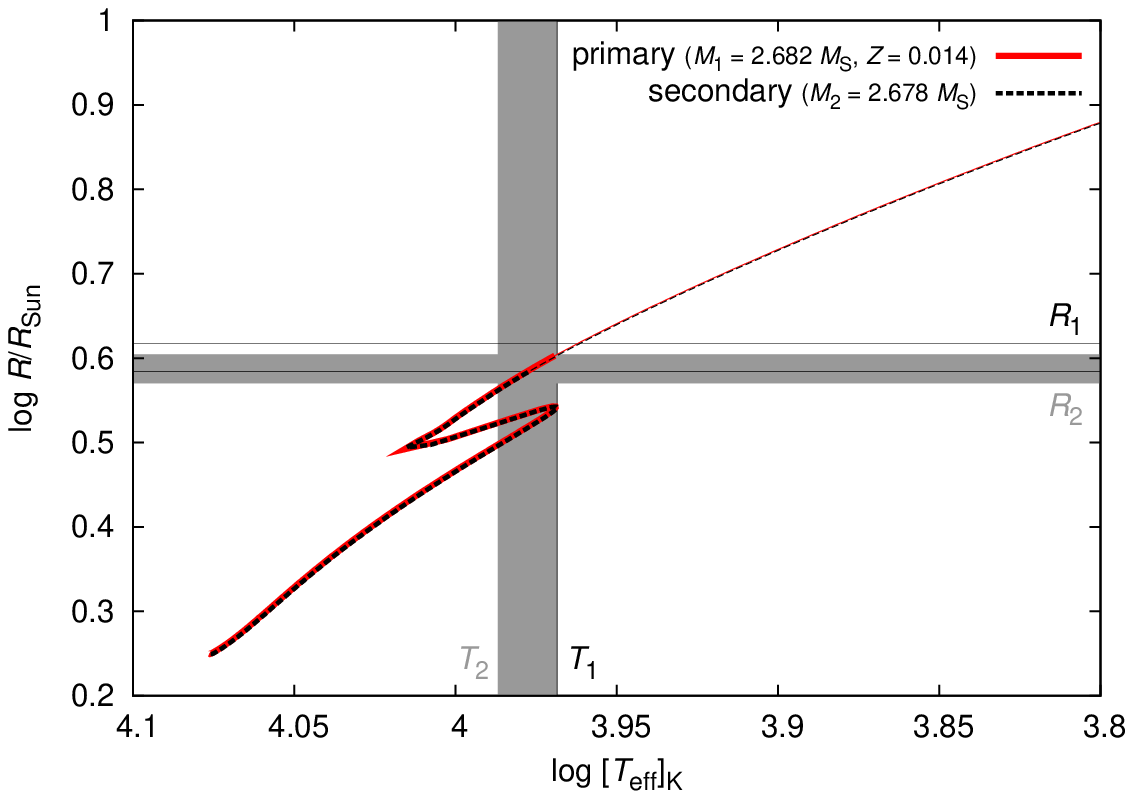}
\caption{Left panels: A Hertzsprung--Russel diagram (top), and
the \tef vs. radius diagram (bottom) showing the stellar evolution
of the primary and secondary components of \ve.
It was computed with the {\tt MESAstar} module,
for the masses $M_1 = 2.62\,M_\odot$, $M_2 = 2.74\,M_\odot$ and
for the metallicity $Z = 0.02$.
The evolutionary tracks are plotted by thick lines from ZAMS up to the age $3.877\times 10^8\,{\rm y}$.
The ranges in temperatures $T_1$, $T_2$, luminosities $L_1$, $L_2$ and
radii $R_1$, $R_2$ inferred from photometry/spectroscopy are denoted by
lines (refer to Table~\ref{fotrvsol}).
There is a strong disagreement between the observations
and the stellar evolution, especially for the primary.
Right panel: same for the masses $M_1 = 2.682\,M_\odot$, $M_2 = 2.678\,M_\odot$
(i.e., the mass ratio very close to one),
and a different value of the metallicity $Z = 0.014$.
The thick lines are terminated at the age $3.938\times 10^8\,{\rm y}$.}
\label{modelhrd}
\end{figure*}

We first calculated the model evolution for the masses
$M_1 = 2.62\,M_\odot$ and $M_2 = 2.74\,M_\odot$ which follow from the free
\phoebe solution with the lowest $\chi^2$ for the combined photometric
and RV data (Table~\ref{fotrvsol}, left). We assumed the same metallicities
of $Z = 0.02$ for both components, of course, the helium abundance
$Y = 0.28$ and the mixing-length parameter $\alpha = 2.0$.
The result is compared to the observed binary properties in
the Hertzsprung--Russell diagram, and the \tef vs. radius diagram
(Figure~\ref{modelhrd}, left panels).
Even though there are uncertainties in temperatures,
luminosities and masses of the individual components
(refer to Table~\ref{fotrvsol}),
their {\em differences\/} are established much more accurately;
e.g., the difference in temperatures $T_2-T_1 \simeq 200\,{\rm K}$
is always present in \phoebe solutions since this is enforced
by the observed light curve and colour indices.

There is a clear disagreement between the photometric/spectroscopic
observations and the predicted stellar evolution in this case.
Since the mass ratio~$q = M_1/M_2 \doteq 0.96$ differs significantly from~1,
the calculated luminosities of the components are {\em always\/} very
different owing to a strong dependence of the stellar
evolution on the mass ($\log L_2/L_\odot$ reaches $\simeq 2.0$ and
$\log L_1/L_\odot \simeq 1.9$), while the observed luminosities are rather
similar ($\log L_1/L_\odot \simeq \log L_2/L_\odot \simeq 2.03$).
A change in neither metallicity nor in the mixing-length parameter could alter
this result since a different value of~$Z$ would shift both tracks in the same
direction, and varying $\alpha$ from 1.5 to 2.5 does not alter evolutionary
tracks significantly before the red giant branch is reached.

As a second test, we took the mass ratio $q$ close to~1,
which is still compatible with the photometric/spectroscopic observations
from a statistical point of view (Table~\ref{fotrvsol}, right).
Because the stellar evolution is very sensitive to the stellar mass,
we may actually use this approach to constrain the mass ratio of V2368~Oph.
The most sensitive indicator seems to be the temperature --
there is approximately a 200\,K difference between $T_1$ and $T_2$,
which corresponds to a $0.004$ to $0.008\,M_\odot$ difference between
$M_1$ and $M_2$, according to our tests.
If we use $M_1 = 2.682\,M_\odot$ and $M_2 = 2.678\,M_\odot$
we also have to decrease the metallicity to $Z = 0.014$,
which shifts both the evolutionary tracks towards higher $T$ and $L$,
in order to match the observed state of V2368~Oph
(Figure~\ref{modelhrd}, right panels).
Another possibility would be to slightly increase the masses
to $M_1 = 2.760\,M_\odot$ and $M_2 = 2.752\,M_\odot$ and to retain
the $Z = 0.02$ value.
To conclude, it is possible to find a consistent solution
for the available photometry and spectroscopy and the stellar evolution,
even though the parameters presented above cannot be considered
as final, because the total mass of the system is not yet constrained
precisely enough.

From the standpoint of stellar evolution, \vo is a very interesting
evolved system with both components leaving the main sequence.
It is in a rapid phase of evolution and consequently may serve
as a very sensitive test case for the stellar-evolution programmes,
provided new, accurate RVs and photometric observations are acquired.
Considering the relatively short distance of the binary from us,
its interferometry would also be of utmost importance,
providing the angular separation
of the components and orbital inclination, consequently a much more
precise parallax and the total mass of the system.
Interferometry can also provide independent constraints on
the component radii.


\section{A comparison with synthetic spectra and the rotation
of the binary components}

\begin{figure*}
\centering
\includegraphics[width=8cm]{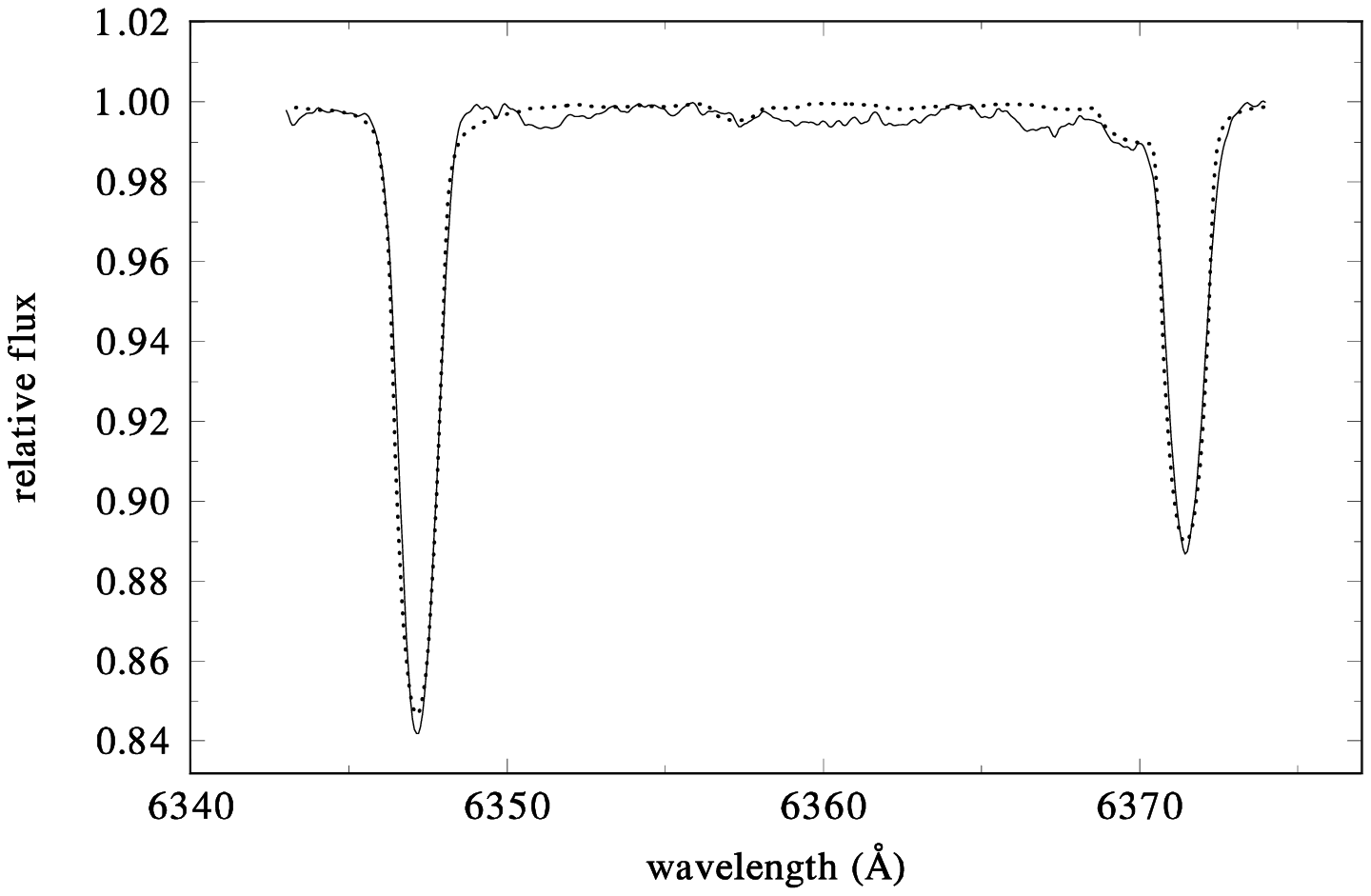}
\includegraphics[width=8cm]{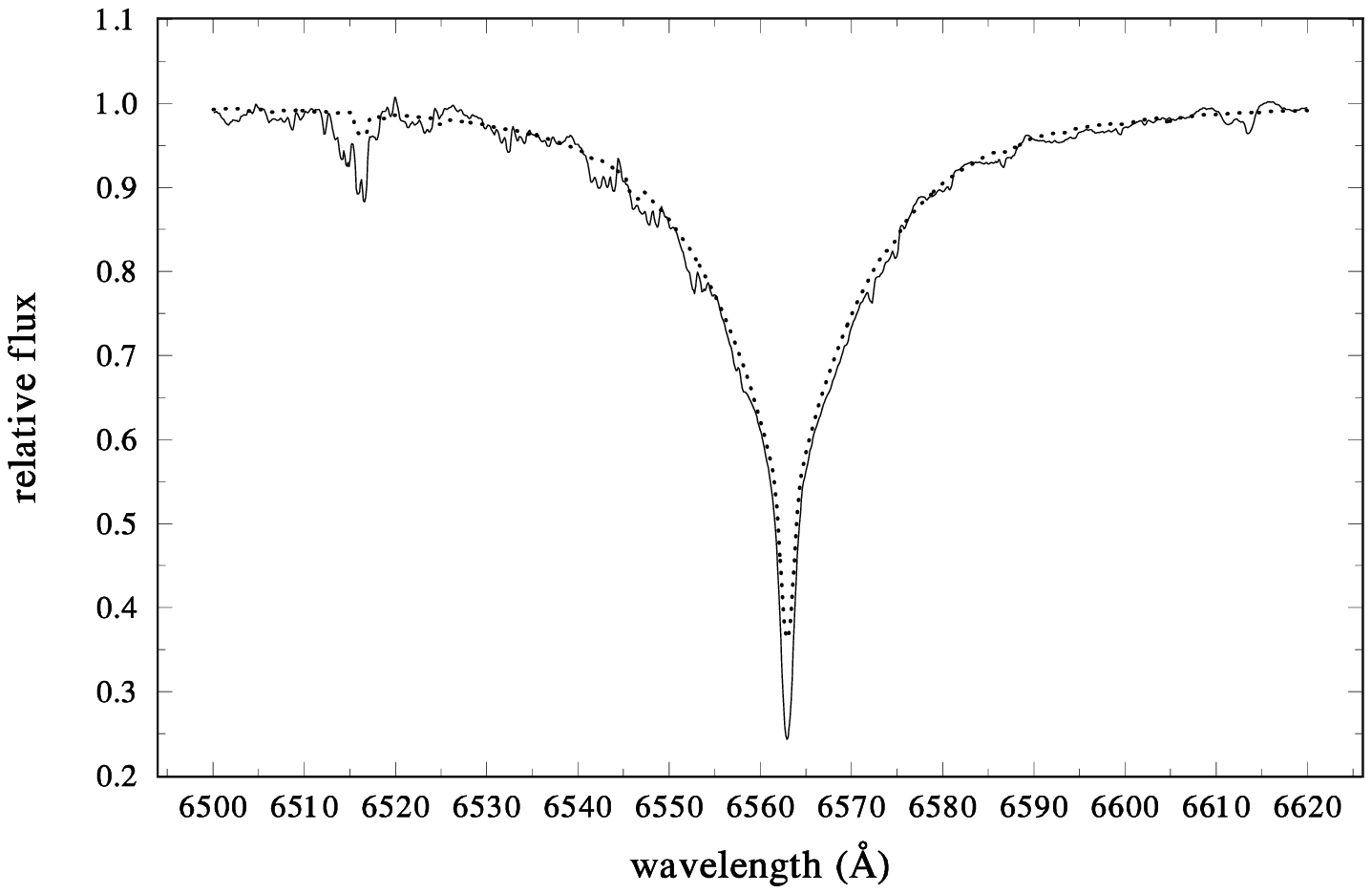} \\
\includegraphics[width=8cm]{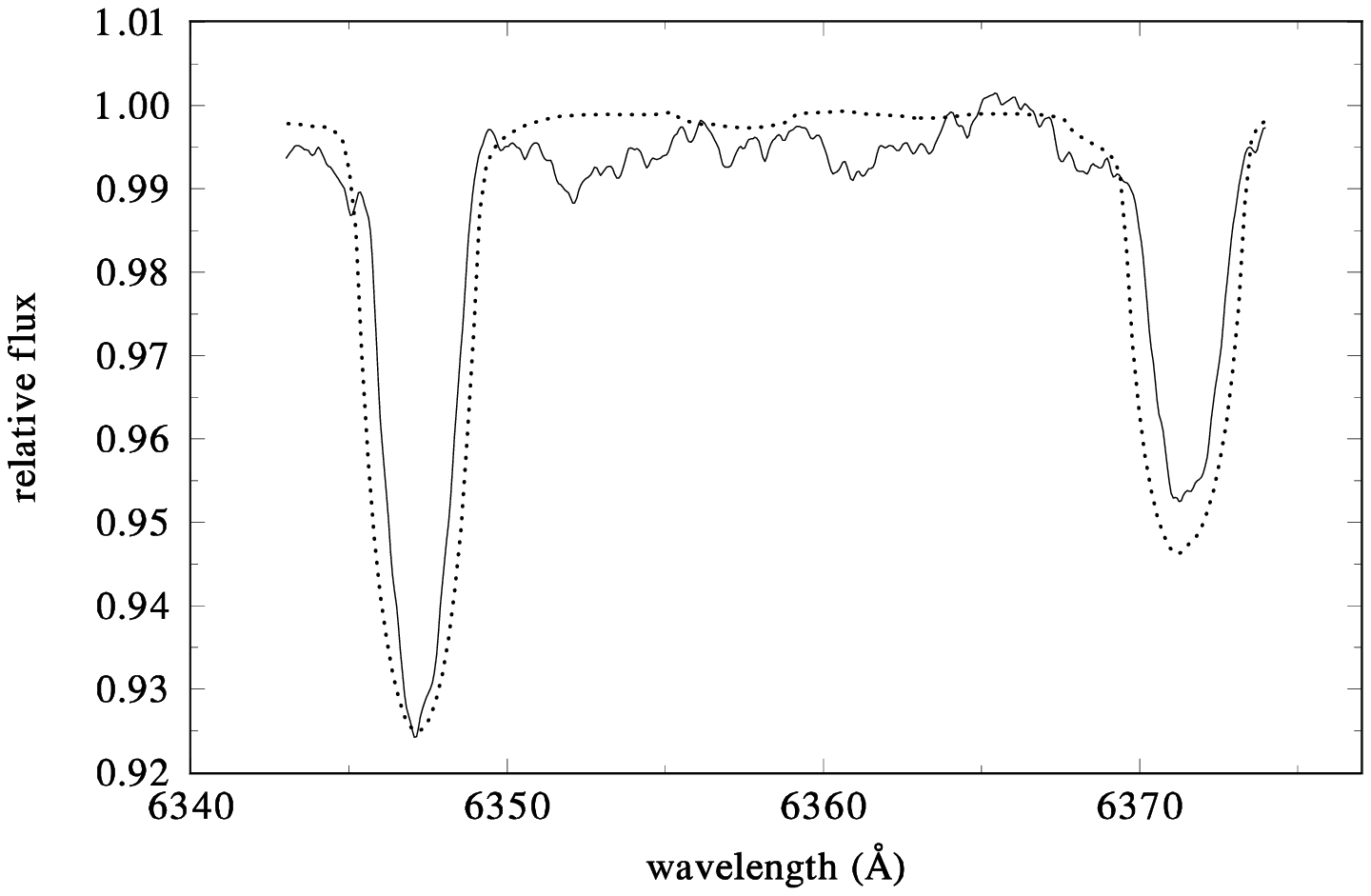}
\includegraphics[width=8cm]{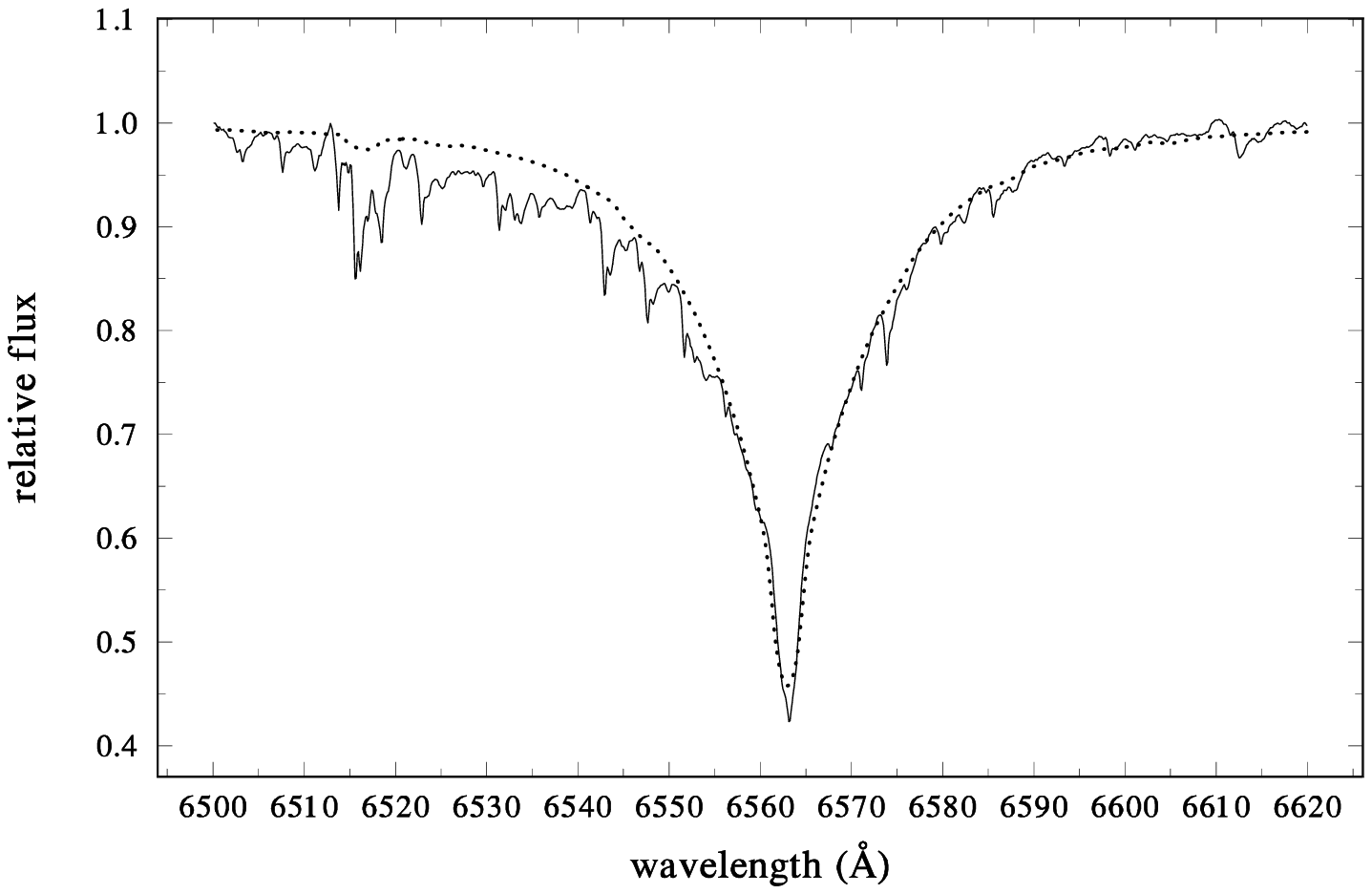}
\caption{A comparison of {\em disentangled\/} (solid lines)
and synthetic (dotted lines) profiles
of the \ion{Si}{II} line (left panels) and \ha line (right)
for the primary (top panels) and secondary components (bottom).
See the text for details.}
\label{sikor}
\label{hakor}
\end{figure*}

As another consistency check, we disentangled the \ion{Si}{II} and
\ha line profiles with the help of the \korel program
\citep{korel1,korel2,korel3,korel4}, keeping the orbital parameters from
the second \phoebe solution fixed, but using the mass ratio
of 0.998 (considering the discussion above). In Fig.~\ref{sikor},
the disentangled line profiles, normalized to their individual
continua using the relative luminosities derived by \phoebee, are
compared with the synthetic line profiles from the Ond\v{r}ejov
library of synthetic spectra prepared and freely distributed
by Dr.~J.~Kub\'at --- see, e.g. \citet{sef1} for details.
We used the synthetic spectra for the parameters close to
the \phoebe results, namely \tef = 9500\,K and \lgg = 3.5, rotationally
broadened in \spefo\ to 40\,\kms for the primary and to 90\,\kms
for the secondary. Varying these values for more than $\pm 5$~\kms
would result in a significant disagreement in the depths and widths
between the observed and synthetic line profiles.

The agreement between the observed and synthetic spectra is satisfactory,
lending some credibility to our result. We warn, however, that the
heterogeneity of our material means that there can still be
rather large uncertainties in the derived masses, radii, and luminosities.
Another study based
on rich and homogeneous observational material is therefore desirable.

Taken at face value, the observed projected  rotational velocities
and the \phoebe solution would imply the stellar rotational periods
of 5\fd0 and 2\fd2 for the primary and secondary, respectively, while
the spin-orbit synchronization in periastron would imply
a rotational period of 10\fd5. It is notable that in their detailed study of
another evolved A-type binary with eccentric orbit, $\theta^2$~Tau,
\citet{tor2011} also found the projected rotational velocity
for the secondary roughly twice as high as for the primary.

In passing, we wish to mention that it is also possible to make
a theoretical prediction of the internal structure
constant based on the current evolutionary models of \citet{claret2004} with
the standard chemical composition of $(X,Z) = (0.70, 0.02)$:
$\log k_2 = -2.498$ for both binary components. Taking the values of
the eccentricity and fractional radii from Table~\ref{fotrvsol} into account,
we can predict a very slow apsidal-motion rate
of $\dot\omega_{\rm obs} = 0.00028$~deg/cycle, which is only 0.27~deg/century.
The relativistic contribution to the apsidal motion is substantial:
$\dot\omega_{\rm rel} = 0.00020$~deg/cycle or about 70\% of the total
apsidal-advance rate \citep{gim85}. In other words, there is little chance
of detecting a measurable apsidal motion for this binary in the foreseeable
future.


\begin{acknowledgements}
We acknowledge the use of the latest version of the \phoebe\ program, developed
and freely distributed by Dr.~A.~Pr\v{s}a; of the latest publicly available
versions of programs \fotel and \korel written and distributed by
Dr.~P.~Hadrava; of the freely distributed {\tt MESAstar} module for stellar
evolution by Dr. B.~Paxton and collaborators; and of the program {\tt UVBYBETA}
written by Dr. T.T.~Moon
and modified by Dr. R.~Napiwotzki.  We also profited from the use of the
library of synthetic spectra prepared and provided by Dr.~J.~Kub\'at.
The constructive criticism of an anonymous referee helped us improve the
presentation of the results.
The research of the Czech authors was supported by the grant
205/06/0304, 205/08/H005, and P209/10/0715 of the Czech Science Foundation
and also from the Research Program MSM0021620860
{\sl Physical study of objects and processes in the solar system
and in astrophysics} of the Ministry of Education of the Czech Republic.
We acknowledge the use of the electronic database from the CDS Strasbourg and
the electronic bibliography maintained by the NASA/ADS system.
\end{acknowledgements}

\bibliographystyle{aa}
\bibliography{citace}


\Online
\begin{appendix}
\section{Photometry}\label{aone}

\begin{table*}
\caption[]{Journal of available photometry of V2368 Oph}\label{journalphot}
\begin{flushleft}
\begin{tabular}{rcrccl}
\hline\hline\noalign{\smallskip}
Station&Time interval& No. of &Passbands&HD of comparison&\ \ \ \ Source\\
       &(HJD$-$2400000)&obs.  &  &/ check star\\
\noalign{\smallskip}\hline\noalign{\smallskip}
 1&54273.4--54977.6& 423&\ubv & 154660$/$154895 & this paper\\
30&52065.8--54276.8& 152&\ubv &       *)     & this paper\\
66&52765.4--52766.6&  10&\ubv & 154660$/$154895, 154445 & this paper\\
61&47912.6--49061.9& 111&$V$  & all-sky      & \protect{\cite{esa97}}\\
93&53055.9--53290.5&  47&$V$  & all-sky      & \protect{\cite{pojm2002}}\\
\noalign{\smallskip}\hline
\end{tabular}\\
\smallskip
{\scriptsize Individual observing stations are distinguished by
running numbers they have in the Prague/Zagreb photometric
archives --- see column {\sl ``Station":}
01~\dots~Hvar 0.65-m, Cassegrain reflector, EMI9789QB tube;
30~\dots~San Pedro M\'artir, 0.84-m reflector, Cuenta-pulsos photometer;
61~\dots~Hipparcos all-sky $H_{\rm p}$ photometry transformed to Johnson $V$;
66~\dots~TNO 0.40-m Cassegrain reflector, SSP5A photometer;
93~\dots~ASAS data archive \protect{\citep{pojm2002}}.\\
\**) All-sky photometry or differential photometry relative to
various comparisons during the first season when \vo was used
as a comparison for observations of U~Oph, then relative to HD~154660 ---
see the text for details.}
\end{flushleft}
\end{table*}
Here we provide detailed comments on the photometric observations used
in this study and the way we treated them.
\begin{enumerate}
\item Hvar observations were secured in 2007, 2008, and 2009 and reduced and
      transformed into the standard \ubv\ system with the HEC22~release
      16.1 reduction program via nonlinear transformation formul\ae\
      \citep{hhj94,hechor98}. This recent version of the program allows
      modelling of variable extinction during the observing night, which
      significantly improves the accuracy of the observations. The
      typical rms errors of the multinight fit to all standard stars used
      to define the transformation formul\ae\ in a given observating season
      are 0\m008 for $V$ and $B$, and 0\m010 for $U$. This is similar
      for observations from other observing stations reduced with the help of
      HEC22.

\item San Pedro M\'artir observations were collected during
      the observational runs
      in 2001--2003 and 2007. Observations were reduced and
      transformed into the standard \ubv\ system with the HEC22~release
      14.1 reduction program via nonlinear transformation formul\ae\
      \citep{hhj94, hechor98}.
      At that station  \vo was used in 2001 as a recommended
      comparison star for the eclipsing binary U~Oph and its
      magnitude difference relative to several comparison
      stars (HD~183324, HD~187458, HD~161132, HD~153808, and HD~144206)
      was derived.
      For all these stars, save HD~183324, the magnitudes and colours
      are well established from the calibrated Hvar all-sky photometry.
      HD~183324=V1431~Aql was found to be a small-amplitude
      $\lambda$ Bootis variable \citep{kuspauwei94}.
      It served for a long time as a comparison star for observations
      of V923~Aql and V1294~Aql in the Photometry of the Bright Northern
      Be Star Programme \citep{hecetal82,hhj94,hvar5,hvar5d} and
      its variability on longer time scales is
      safely excluded by numerous Hvar observations. The mean all-sky \ubv\
      magnitudes of HD~183324 are accurately derived. It was actually
      used as a comparison for \vo only on the night JD~2452065,
      and we feel that its 2~mmag microvariability is not critical
      for the purpose of this study.
      For 13 observations secured on JD~2452071 (when we recorded
      the first eclipse of \ve), it was not
      possible to derive the differential values for it so we
      adopted its all--sky values instead, since enough standard stars
      had been observed during the night, and the nightly transformation
      coefficients (extinction and its variations and the zero points)
      could be derived. As soon as we realized that \vo is a variable,
      its subsequent observations in 2002 and 2003 were carried out
      differentially, relative to HD~154660=HR~6361. This A9V star is a visual
      binary ADS~10347A with a close companion ADS~10347B at 20\farcs3,
      which is some 3\m35 fainter than ADS~10347A.
      The 2007 observations were obtained with a larger diaphragm so that
      the light of the visual component ADS~10347B was recorded
      with the brighter component ADS~10347A = HD~6361. We carried out
      dedicated observations at Hvar to derive the total magnitude of
      both visual components and added this value to the magnitude
      differences var -- comp. from this season. In all other instances,
      observations were carried out in such a way as to keep ADS~10347B
      outside the diaphragm.

\item Hipparcos all-sky $H_p$ broad--band magnitudes secured
      between 1989 and 1993 (\cite{esa97}) were transformed to the standard
      Johnson $V$ magnitude with the nonlinear transformation formula
      derived by \cite{hec98}. The rms error of the fit per
      1 observations is 0\m0067. For the solutions, the transmission and
      the limb darkening coefficients for the \hp\
      passband were considered, however. All data with error flags larger
      than 1 and one deviating point at HJD~2448661.4682 were omitted.

\item TNO (Tubitak National Observatory) observations were secured during
      two consecutive nights in 2003
      and were reduced and transformed into the standard \ubv\ system
      with the HEC22~release 14.1 reduction program via nonlinear
      transformation formul\ae\ \citep{hhj94,hechor98}.

\item ASAS $V$ magnitude observations were extracted from the public
ASAS database \citep{pojm2002}; we used the data from the diaphragm,
which gives the smallest rms errors and omitted a few clearly deviating
data points.
\end{enumerate}

The journal of all photometric observations is in Table~\ref{journalphot}.
Homogenized \ubv\ magnitudes of all comparison and check stars used
can be found in Table~\ref{compcheck}.

\begin{table}
\caption[]{Comparisons and check stars of \ve}\label{compcheck}
\begin{flushleft}
\begin{tabular}{rllrr}
\hline\hline\noalign{\smallskip}
 HD/BD & Other ident.& \ \ \ $V$  &   \bv &    \ub \\
\noalign{\smallskip}\hline\noalign{\smallskip}
 154660  &    HR 6361     &  6.357   &    0.211     &   0.103   \\
$-01^\circ$ 3292B&    ADS 10347B  &  9.71    &    0.66      &   0.15    \\
         &    ADS 10347AB &  6.308   &    0.227     &   0.094   \\
 154895  &    HR~6367     &  6.058   &    0.075     &   0.028   \\
 183324  &    $V1431$~Aql &  5.801   &    0.086     &   0.067   \\
 187458  &    HR~7550     &  6.660   &    0.426     & $-0.056$  \\
 162132  &    HR~6641     &  6.494   &    0.085     &   0.075   \\
 153808  &    $\epsilon$ Her     &  3.916   &   $-0.024$     &  $-0.088$   \\
 144206  &    $\upsilon$ Her     &  4.739   &   $-0.096$     &  $-0.326$   \\
\noalign{\smallskip}\hline
\end{tabular}\\
\smallskip
{\scriptsize Magnitude and colours of ADS 10347AB are values
resulting from co-added flux of HR 6361 and BD $-01^\circ$ 3292B
measured simultaneously though a larger diaphragm in the photometer.}\\
\end{flushleft}
\end{table}


\section{Spectroscopy}\label{btwo}

\begin{table}
\caption[]{Journal of spectroscopic data of V2368 Oph}
\label{journalspect}
\begin{flushleft}
\begin{tabular}{cccll}
\hline\hline\noalign{\smallskip}
Spg.& Time interval& No. of&\ \ \ \ \ \ \ \ Station, telescope\\
no. & (HJD$-$2400000)&spectra &\ \ \ \ \ \ \ \ and instrument& \\
\noalign{\smallskip}\hline\noalign{\smallskip}
A &54266.4--54638.4& 11&OND 2.0-m, grating spg.\\
B &54339.7--55102.6& 19&DAO 1.22-m, grating spg.\\
C &52745.8--52749.0& 21&SPM 2.1-m, echelle spg.\\
D &54193.9--54200.0& 21&SPM 2.1-m, echelle spg.\\
E &54250.8--54253.0& 12&SPM 2.1-m, echelle spg.\\
\noalign{\smallskip}\hline
\end{tabular}\\
\smallskip
\end{flushleft}
\end{table}

The journal of all spectroscopic observations can be found in
Table~\ref{journalspect}.
The individual data files are identified there by letters.
Below, we provide a few comments on them.

\begin{itemize}
\item {\sl File A} CCD spectra of V2368 Oph covering the
wavelength region 6260--6750~\ANG.
They were secured with a SITe-005 $800\times2000$ CCD detector
attached to the medium 0.7-m camera of the coud\'{e} focus of
the Ond\v{r}ejov 2.0~m telescope (OND).
The spectra were obtained between June 2007 and June 2008
and have a linear dispersion of 17.2~\Ame (red) and
a~2-pixel resolving power of about 12600 (11-12~\kms per pixel).
Their $S/N$ ranges from 50 (1~spectrum) to 370, and the majority have
$S/N$ over 200.
\item{\sl File B} CCD spectra covering the wavelength region
6150--6750~\ANG\ with a resolution of 6.6~\kms per pixel.
They were obtained at the DAO 1.22-m telescope between August 2007 and
September 2009 and have a reciprocal linear dispersion of 10~\Ame. The
detector used was a SITe-4 $4096 \times 2048$ CCD, and the 3-pixel
resolving power was about 15000. Their $S/N$ ranges from 100 to 370.
\item{\sl File C} CCD echelle spectra secured with the 2.14-m reflector
of the SPM observatory in April 2003. The CCD detector has $1024 \times 1024$
pixels, and the setting used covered the wavelength region from 4000 to
6700~\ANG\ in grating orders 33 to 60. The nominal resolution of the
spectrograph is 18000 at 5000~\ANG, which translates into 2-pixel
resolution of about 17 \ks.
\item{\sl File D} Another set of echelle CCD spectra from SPM,
 secured in April 2007
\item{\sl File E} The third set of echelle CCD spectra from SPM,
 secured from May 30 to June 1, 2007. The $S/N$ of the SPM spectra
ranges between 120 and 500 for the red, and 85 to 230 for the blue
parts of the spectra.
\end{itemize}

The initial reductions of the DAO spectra (bias subtraction, flatfielding
and conversion to 1-D images) were carried out by SY in IRAF. The initial
reduction of the SPM and OND spectra was carried out by PE and by
Dr. M.~\v{S}lechta, respectively, including the wavelength calibration.
The remaining reductions of all spectra (including wavelength calibration for
the DAO spectra, continuum rectification, and removal of cosmics and flaws)
was carried out with the program \spefo \citep{sef0,spefo}.

\end{appendix}
\end{document}